\documentclass{elsart}
\usepackage{graphicx}
\usepackage{epsfig}
\usepackage{axodraw}
\usepackage{amssymb}

\newcommand{\Li}{\mathop{\mathrm{Li}}\nolimits}
\newcommand{\Si}{\mathop{\mathrm{S}}\nolimits}
\newcommand{\trg}[3]{
\begin{picture}(35,30)(5,13)
\Line(5,5)(20,30)
\Line(35,5)(20,30)
\Line(5,5)(35,5)
\Vertex(5,5){2}
\Vertex(35,5){2}
\Vertex(20,30){2}
\Text(20,0)[]{$\scriptstyle #1$}
\Text(13,25)[r]{$\scriptstyle #2$}
\Text(27,25)[l]{$\scriptstyle #3$}
\end{picture}
}

\begin{document}

\begin{frontmatter}

\title{Calculating four-loop tadpoles with one non-zero mass}

\author{Bernd A. Kniehl},
\ead{kniehl@desy.de}
\author{Anatoly V. Kotikov\thanksref{label1}}
\ead{kotikov@theor.jinr.ru}
\address{II. Institut f\"ur Theoretische Physik, Universit\"at Hamburg,
Luruper Chaussee 149, 22761 Hamburg, Germany}
\thanks[label1]{Permanent address: Bogoliubov Laboratory of Theoretical
Physics, Joint Institute for Nuclear Research, 141980, Dubna (Moscow region),
Russia.}

\begin{abstract}
An efficient method to calculate tadpole diagrams is proposed.
Its capability is demonstrated by analytically evaluating
two four-loop tadpole diagrams of current interest in the literature,
including their $O(\epsilon)$ terms in $D=4-2\epsilon$ space-time dimensions.
\end{abstract}

\end{frontmatter}

\section{Introduction}

Calculations of higher-order corrections are very important for precision
tests of the Standard Model in present and future high-energy-physics
experiments.  
The complexity of such calculations and the final results strongly increases
with the number of quantum loops considered, and one rapidly reaches the
limits of the present realm of possibility when one attempts to exactly
account for all mass scales of a given problem, already at the two-loop level.
To simplify the calculations and also the final expressions, various types of
expansions were proposed during the last couple of years
(see Ref.~\cite{AS1,AS2,DaTa} and references cited therein).
These approaches usually provide a possibility to reduce the problem of
evaluating complicated Feynman integrals to the calculation of tadpoles
$T_{0,m}(\alpha_1,\alpha_2)$
and loops with massless propagators $L_q(\alpha_1,\alpha_2)$,
defined in Eqs.~(\ref{eq:one}) and (\ref{eq:two}), respectively, which have
simple representations.

As a first step, let us to introduce some definitions to be used below.
All the calculations are performed in Euclidean momentum space of dimension
$D=4-2\epsilon$ using dimensional regularisation with 't~Hooft mass scale
$\mu$.
For the Euclidean integral measure, we use the short-hand notation
$Dk\equiv {\mu}^{4-D}d^Dk/{\pi}^{D/2}$.
The dotted and solid lines of any diagram correspond to massless and massive
Euclidean propagators, represented graphically as
\begin{equation}
\frac{1}{(k^2)^\alpha}=\hspace{3mm}
\mbox{{
\begin{picture}(30,10)(5,4)
\Gluon(0,5)(30,5){1}{5}
\Vertex(0,5){1}
\Vertex(30,5){1}
\Text(15,11)[b]{$\scriptstyle\alpha$} 
\end{picture}
}}\hspace{1mm}
,\qquad \frac{1}{(k^2+m^2)^\alpha}=\hspace{3mm}
\mbox{{
\begin{picture}(30,10)(5,4)
\Line(0,5)(30,5)
\Vertex(0,5){1}
\Vertex(30,5){1}
\Text(15,11)[b]{$\scriptstyle\alpha$} 
\Text(15,3)[t]{$\scriptstyle m^2$}
\end{picture}
}}\hspace{1mm},
\end{equation}
where $\alpha$ and $m$ denote the index and mass of the considered line,
respectively.
Unless stated otherwise, all solid lines have the same mass $m$.
Lines with index 1 and mass $m$ are not marked.

Let us now discuss the rules used in our calculation.
Firstly, massive tadpoles $ T_{0,m}(\alpha_1,\alpha_2)$ are integrated
according to the identity
\begin{eqnarray}
T_{0,m}(\alpha_1,\alpha_2)&\equiv&
\int\frac{Dk}{(k^2)^{\alpha_1}{(k^2+m^2)}^{\alpha_2}}\equiv\hspace{3mm}
\mbox{{
\begin{picture}(50,30)(0,13)
\GlueArc(20,15)(15,90,270){1}{6}
\CArc(20,15)(15,-90,90)
\Vertex(20,30){2}
\Vertex(20,0){2}
\Line(20,0)(17,-5)
\Line(20,0)(23,-5)
\Text(3,18)[r]{$\scriptstyle\alpha_1$} 
\Text(38,18)[l]{$\scriptstyle\alpha_2$} 
\end{picture}
}}
\nonumber\\
&=&\frac{\displaystyle R(\alpha_1,\alpha_2)}
        {\displaystyle (m^2)^{\alpha_1+\alpha_2-D/2}},
\label{eq:one}
\end{eqnarray}
where
\begin{equation}
R(\alpha_1,\alpha_2)=
\frac{\Gamma(D/2-\alpha_1)\Gamma(\alpha_1 +\alpha_2 -D/2)}{\Gamma(\alpha_2)
\Gamma(D/2)}
\end{equation}
and $\Gamma$ is Euler's Gamma function.
Secondly, massless loops $ L_q(\alpha_1,\alpha_2)$ with the external 
momentum $q$ are integrated according to the identity
\begin{eqnarray}
L_q(\alpha_1,\alpha_2)
&\equiv&\int\frac{Dk}{(k^2)^{\alpha_1}{(k-q)}^{\alpha_2}}  
\equiv\hspace{3mm}
\mbox{{
\begin{picture}(60,30)(0,13)
\GlueArc(30,0)(30,25,150){1}{6}
\GlueArc(30,30)(30,-25,-150){1}{6}
\Line(5,15)(-5,15)
\Line(55,15)(65,15)
\Vertex(5,15){2}
\Vertex(55,15){2}
\Text(30,37)[b]{$\scriptstyle\alpha_1$} 
\Text(0,12)[t]{$q$} 
\Text(30,-5)[t]{$\scriptstyle\alpha_2$} 
\end{picture}
}}
\nonumber\\
&=&A(\alpha_1,\alpha_2)\hspace{3mm}
\mbox{{
\begin{picture}(70,30)(0,4)
\Gluon(5,5)(65,5){1}{6}
\Vertex(5,5){2}
\Vertex(65,5){2}
\Line(5,5)(-5,5)
\Line(65,5)(75,5)
\Text(33,10)[b]{$\scriptstyle\alpha_1+\alpha_2-D/2$} 
\Text(-3,-5)[b]{$q$} 
\end{picture}
}}\hspace{3mm},
\label{eq:two}
\end{eqnarray}
where
\begin{equation}
A(\alpha_1,\alpha_2)= 
\frac{a(\alpha_1) a(\alpha_2)}{a(\alpha_1 +\alpha_2 -D/2)},\qquad
a(\alpha)=\frac{\Gamma(D/2-\alpha)}{\Gamma(\alpha)}.
\end{equation}

Recently, two examples of four-loop tadpoles were calculated in
Ref.~\cite{CKMS}.
Certain $O(\epsilon^0)$ parts of these results, denoted as $N_{10}$ and
$N_{20}$, were presented there only numerically.\footnote{%
During the preparation of this article, we were informed by K.G. Chetyrkin
about a paper \cite{Schroder} which also contains analytic results for
$N_{10}$ and $N_{20}$.
Our results are in full agreement with those of Ref.~\cite{Schroder}.}
Also the $O(\epsilon)$ terms beyond $N_{10}$ and $N_{20}$ are of current
interest in the literature \cite{dec} and so far only known numerically.
In conjunction with the results of Ref.~\cite{KKOV06}, they allow us to derive
analytic results for several four-loop master integrals which are
indispensable for modern calculations at this level of accuracy~\cite{KnKo06}.

The subject of this letter is to advertise a powerful technique to evaluate
such tadpoles analytically.
The simplifications encountered in the examples at hand suggest that this
technique might also be useful for more complicated diagrams.
Therefore, we wish to introduce it to the interested reader.

The content of this letter is as follows.
Section~\ref{sec:two} explains the calculational technique in general.
Sections~\ref{sec:three} and \ref{sec:four} describe its application to the
two integrals $N_{10}$ and $N_{20}$ of Ref.~\cite{CKMS}, respectively, also
including their $O(\epsilon)$ terms.
A summary is given in Section~\ref{sec:five}.

\section{Technique}
\label{sec:two}

The core of the technique is a master formula to represent a loop with two
massive propagators as an integral whose integrand contains a new propagator
with a mass that depends on the variable of integration.
This formula has the following form:
\begin{eqnarray}
\mbox{{
\begin{picture}(60,30)(0,13)
\Curve{(5,15)(30,25)(55,15)}
\Curve{(5,15)(30,5)(55,15)}
\Vertex(5,15){2}
\Vertex(55,15){2}
\Line(5,15)(-5,15)
\Line(55,15)(65,15)
\Text(30,27)[b]{$\scriptstyle\alpha_1,\,m_1$} 
\Text(30,3)[t]{$\scriptstyle\alpha_2,\,m_2$} 
\Text(0,12)[t]{$q$} 
\end{picture}
}}\hspace{3mm}&=&
   \frac{\Gamma(\alpha_1+\alpha_2-D/2)}{\Gamma(\alpha_1)\,\Gamma(\alpha_2)}
\nonumber\\
&&{}\times
\int_0^1 \frac{ds}{(1-s)^{\alpha_1+1-D/2}s^{\alpha_2+1-D/2}}\hspace{3mm}
\mbox{{
\begin{picture}(70,30)(0,4)
\Line(5,5)(65,5)
\Vertex(5,5){2}
\Vertex(65,5){2}
\Line(5,5)(-5,5)
\Line(65,5)(75,5)
\Text(33,10)[b]{$\scriptstyle\alpha_1+\alpha_2-D/2$} 
\Text(33,3)[t]{$\scriptstyle \frac{m_1^2}{1-s}+\frac{m_2^2}{s}$} 
\Text(-3,-5)[b]{$q$} 
\end{picture}
}}
\hspace{3mm}.
\label{eq:master}
\end{eqnarray}
It was introduced in Euclidean and Minkowski spaces in Refs.~\cite{FKVpl} and
\cite{FKVnp}, respectively.
Here, we are working in Euclidean space and will, thus, follow
Ref.~\cite{FKVpl}.
Some recent applications in Minkowski space may be found in Ref.~\cite{KKOV}.
Special cases of Eq.~(\ref{eq:master}) were considered in
Ref.~\cite{DEM}, where a differential-equation method
\cite{DEM,DEMRem} was introduced.
The latter only requires that quite simple diagrams are calculated directly.
The results for more complicated diagrams may be reconstructed by
integrating the results for simpler diagrams over external parameters.
In many cases, the results for complicated diagrams may be obtained by
integrating the one-loop tadpoles of Eq.~(\ref{eq:one}). 
The method is now very popular for the calculation of complicated Feynman
integrals; for recent articles, see Ref.~\cite{papers} and references cited
therein.

Here, we shall follow a similar strategy.
Applying Eq.~(\ref{eq:master}), we shall represent the results of
Ref.~\cite{CKMS} as integrals over one-loop tadpoles, which in turn contain
propagators with masses that depend on the variable of integration. 
The case of $N_{10}$ will be considered in detail.

We would like to note that Eq.~(\ref{eq:master}) can be applied successfully
to diagrams containing one-loop self-energy sub-diagrams.
Such sub-diagrams are frequently generated by the application of the
integration-by-parts technique \cite{byparts}.
Such an application \cite{DEM,DEMRem} leads to a differential equation for the
original diagram (in exceptional cases, the equation degenerates to an
algebraic relation) with an inhomogeneous term that depends on less 
complicated diagrams usually containing one-loop self-energy sub-diagrams
(see, for example, Refs.~\cite{FKVpl,FKVnp}).

\section{First example}
\label{sec:three}

As the first example, we consider the case of $N_{10}$ in Ref.~\cite{CKMS},
which may be represented graphically as\footnote{The graphic representations
of $N_{10}$ and $N_{20}$ are adopted from Ref.~\cite{CKMS}.
A propagator with a point is equal to one with the index 2.} 
\begin{equation}
m^2
\raisebox{-0.9cm}{{
\includegraphics[height=2.0cm,bb=71 115 697 720]{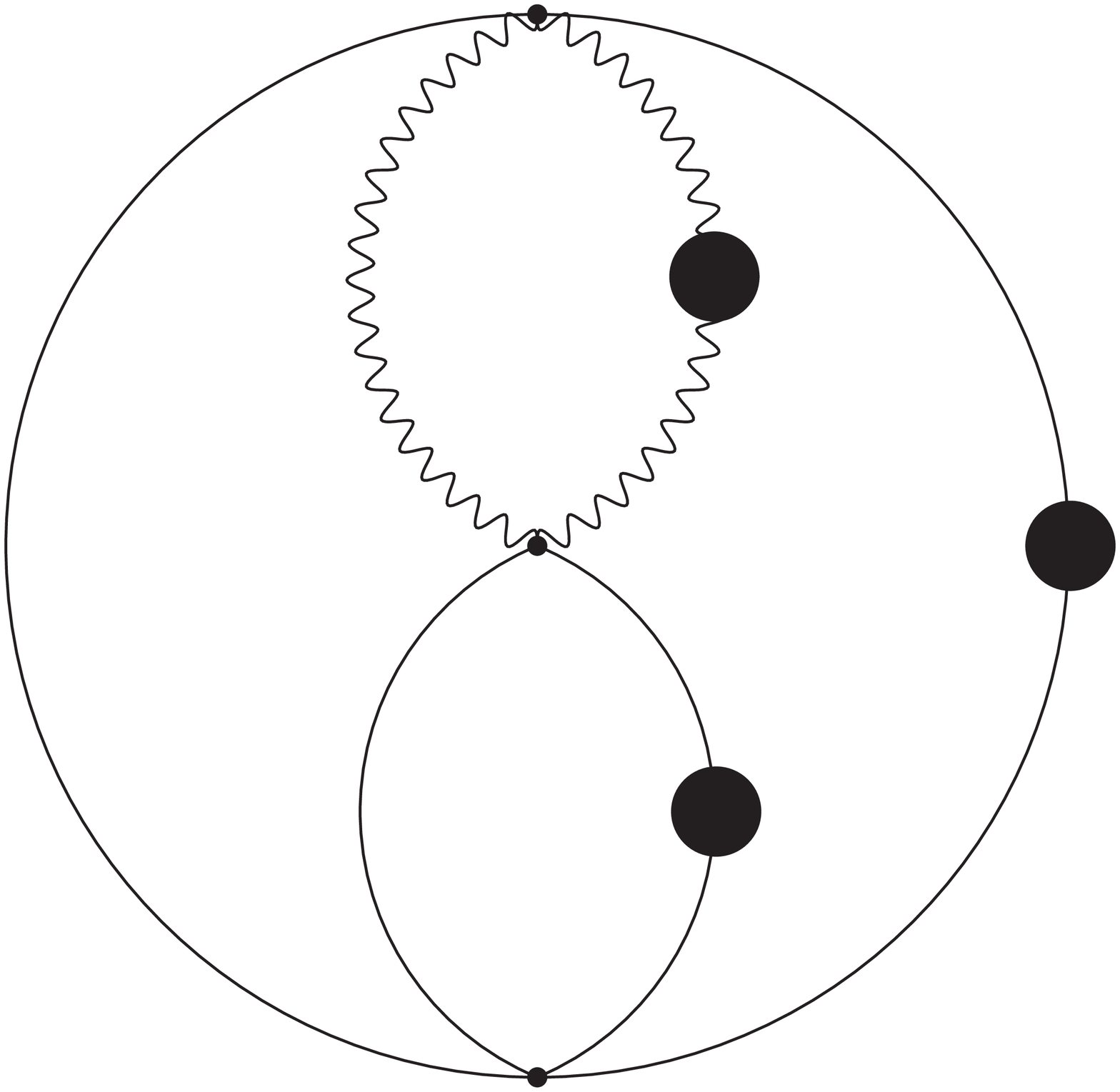}
}}
+m^{2(1-\epsilon)}\Gamma(\epsilon)
\raisebox{-0.9cm}{{
\includegraphics[height=2.0cm,bb=71 114 697 720]{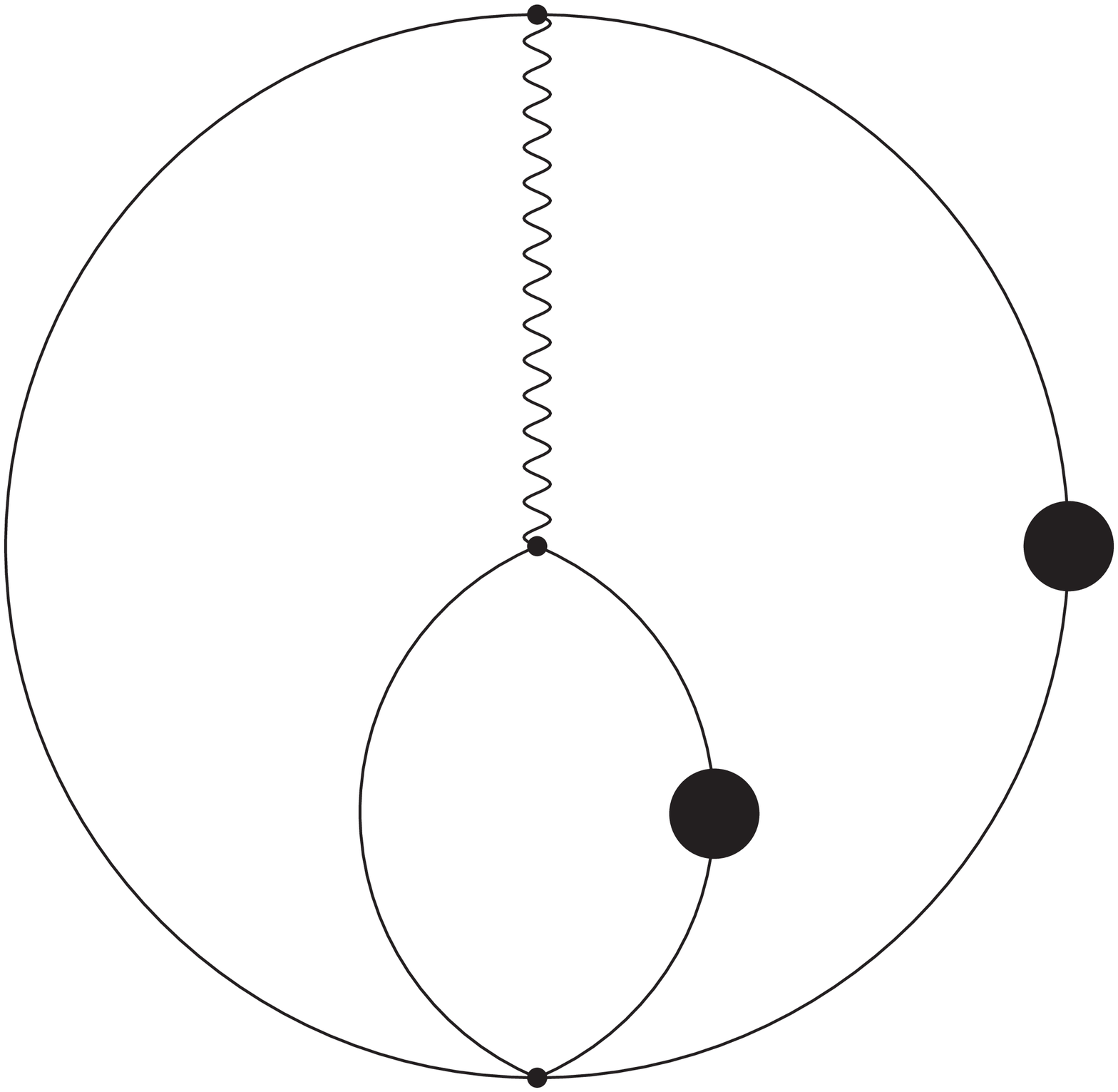}
}}
= \frac{\Gamma^4(1+\epsilon)}{(m^2)^{4\epsilon}}
N_{10}(\epsilon),
\label{NM1}
\end{equation}
where $N_{10}(\epsilon)=[N_{10}+M_{10}\epsilon+O(\epsilon^2)]/(1-\epsilon)$.
For $\epsilon=0$, we recover $N_{10}=N_{10}(0)$.
Applying Eq.~(\ref{eq:two}) to the massless loop in the left diagram and
Eq.~(\ref{eq:master}) with $\alpha_1=1$ and $\alpha_2=2$ to the interior
massive loops and to the loops with exterior propagators, we easily obtain the
following representation for $N_{10}(\epsilon)$:
\begin{equation}
N_{10}(\epsilon) ={m^2\over\epsilon}\left[\frac{1}{m^{2\epsilon}}
J_1(0)-\frac{\Gamma^2(1-\epsilon)}{\Gamma(1-2\epsilon)}J_1(1)\right],
\end{equation}
where
\begin{equation}
J_{1}(a)=
\frac{(1-\epsilon)(m^2)^{4\epsilon}}{\Gamma(1+\epsilon)}
\int_0^1 \frac{ds_1}{s_1 \eta_1^{\epsilon}}
\int_0^1 \frac{ds_2}{s_2 \eta_2^{\epsilon}}
T_{0,M_1,M_2}(1+a\epsilon,1+\epsilon,1+\epsilon),
\end{equation}
$M^2_i=m^2/\eta_i$, $\eta_i=s_i(1-s_i)$, 
and $T_{M_1,M_2,M_3}(\alpha_1,\alpha_2,\alpha_3)$,
where the index $\alpha_i$ belongs to the mass $M_i$, is the one-loop tadpole
involving three massive propagators,
\begin{equation}
T_{M_1,M_2,M_3}(\alpha_1,\alpha_2,\alpha_3)=
\hspace{3mm}
\trg{\alpha_1}{\alpha_2}{\alpha_3}
\hspace{3mm}.
\vspace{5mm}
\end{equation}
Expanding the propagator $(k^2+M^2_2)^{-\alpha_2}$ in 
$T_{0,M_1,M_2}(\alpha_3,\alpha_1,\alpha_2)$ as
\begin{equation}
\frac{1}{(k^2+M^2_2)^{\alpha_2}}=\sum_{n=0}^{\infty}
\frac{\Gamma(n+\alpha_2)}{n!\Gamma(\alpha_2)} \,
\frac{\mu_{12}^n}{(k^2+M^2_1)^{n+1+\epsilon}},
\label{exp}
\end{equation}
where $\mu_{ij}=M^2_i-M^2_j$,
and using Eq.~(\ref{eq:one}) for the tadpole
$T_{0,M_1}(\alpha_1,\alpha_2)$, we obtain
\begin{equation}
T_{0,M_1,M_2}(\alpha_3,\alpha_1,\alpha_2) = 
\frac{R(\alpha_3,\alpha_1+\alpha_2)}{(M_1^2)^{\overline{\alpha} -D/2}}
\,{}_2F_1(\overline{\alpha} -D/2,\alpha_2;\alpha_1+\alpha_2;\overline{x}),
\end{equation}
where $\overline{\alpha}=\alpha_1+\alpha_2+\alpha_3$, $\overline{x}=1-x$,
$x=M^2_2/M^2_1=\eta_1/\eta_2$, and ${}_2F_1$ denotes a hyper-geometric
function  \cite{Gradshteyn}.
Setting $\alpha_i=1+a_i\epsilon$ and expanding the hyper-geometric function, we
have
\begin{eqnarray}
\lefteqn{
T_{0,M_1,M_2}(1+a_3\epsilon,1+a_1\epsilon,1+a_2\epsilon)
=\frac{1}{(M_1^2)^{1+(\overline{a}+1)\epsilon}}\,
\frac{\Gamma(1+\epsilon)}{\overline{x}(1-\epsilon)} 
\left\{-\ln x 
\vphantom{\frac{(a_2+a_3+1)^2}{6}} \right.}
\nonumber\\
&&{}+ \epsilon \left[\frac{a_2+a_3+1}{2} \ln^2 x
-(a_1+a_2) \Li_2(\overline{x}) \right] 
+ \epsilon^2 \left[ \frac{(a_2+a_3+1)^2}{6} \ln^3 x
\right.
\nonumber\\
&&{}-\Bigl((a_3+1)(\overline{a}+1)-1\Bigr)\zeta(2)\ln x
+a_1(a_3+1)\ln x\Li_2(\overline{x})
\nonumber\\
&&{}+\left.\left.\vphantom{\frac{(a_2+a_3+1)^2}{6}}
(a_1+a_2)^2\Li_3(\overline{x})
-\Bigl(a_2(\overline{a}+1)-a_1(a_3+1)\Bigr)\Si_{1,2}(\overline{x})
\right]\right\}
+ O(\epsilon^3),
\end{eqnarray}
where $\overline{a}= a_1+a_2+a_3$ and
\begin{eqnarray}
\Si_{n,m}(x)&=&\frac{(-1)^{n+m-1}}{(n-1)!m!} \int_0^1 \frac{dy}{y} 
\ln^{n-1}(y) \ln^{m}(1-xy),
\nonumber\\
\Li_n(x)&=&\Si_{n-1,1}(x)
\end{eqnarray}
denote the generalised and ordinary poly-logarithms (see, for example,
Ref.~\cite{DD}), respectively.
Thus, we obtain
\begin{eqnarray}
N_{10}(\epsilon)&=&
\int_0^1 \frac{ds_1}{s_1\eta_1^{\epsilon}}
\int_0^1 \frac{ds_2}{s_2\eta_2^{\epsilon}} \, 
\frac{\eta_1}{\overline{x}}
\left\{-\frac{1}{2}\ln^2 x
+\ln x \ln \eta_1 
\right.
\nonumber\\
&&{}+\epsilon \left[\frac{19}{6} \ln^3 x 
+4 \zeta(2)\ln x
-\ln x\Li_2(\overline{x})
- \left(3 \ln^2 x - 2 \Li_2(\overline{x})\right) \ln \eta_1 
\right.
\nonumber\\
&&{}+\left.\left.
\frac{7}{2} \ln x  \ln^2 \eta_1
\right]\right\}
+ O(\epsilon^2).
\end{eqnarray}
Exploiting the $s_1 \leftrightarrow s_2$ symmetry,
we find
\begin{eqnarray}
N_{10}(\epsilon)&=& \frac{1}{8}
\int_0^1 ds_1\int_0^1 
 \frac{ds_2}{\eta_2 - \eta_1}
\left\{\ln x \ln (\eta_1\eta_2) 
+\epsilon \left[ \frac{61}{12} \ln^3 x 
\right.\right.
\nonumber\\
&&{}+8 \zeta(2) \ln x
+\left( \Li_2(\overline{x})-\Li_2\left(-\frac{\overline{x}}{x}\right)\right)
\ln (\eta_1\eta_2)
\nonumber\\
&&{}+\left.\left.
\frac{3}{4}\ln x\ln^2 (\eta_1 \eta_2) 
\right]\right\}
+ O(\epsilon^2).
\end{eqnarray}

We first concentrate on the $O(\epsilon^0)$ term $N_{10}$.
Using the standard replacement $s_i=(1+\xi_i)/2$, we obtain
\begin{eqnarray}
N_{10} &=& \frac{1}{2}\int_0^1d\xi_1
\int_0^1 \frac{d\xi_2}{\xi_1^2-\xi_2^2} 
\left(\ln^2\frac{1-\xi_1^2}{4} - \ln^2\frac{1-\xi_2^2}{4}\right)
\\ 
&=&\frac{1}{2}\int_0^1d\xi_1
\int_0^1 \frac{d\xi_2}{\xi_1^2-\xi_2^2} 
\left\{\left[\ln^2 (1-\xi_1^2) -2\ln 4 \ln (1-\xi_1^2)\right]
- [\xi_1\leftrightarrow\xi_2]\right\}.
\nonumber
\end{eqnarray}
Exploiting the $\xi_1 \leftrightarrow \xi_2$  symmetry, we find
\begin{eqnarray}
N_{10} &=& \int_0^1 d\xi_1
\left[\ln^2 (1-\xi_1^2) -2\ln 4 \ln (1-\xi_1^2) \right]
\int_0^1 \frac{d\xi_2}{\xi_1^2-\xi_2^2}
\nonumber\\ 
&=& - \frac{1}{2}\int_0^1\frac{d\xi_1}{\xi_1} 
\ln\frac{1-\xi_1}{1+\xi_1}
\left[\ln^2 (1-\xi_1^2) -2\ln 4 \ln (1-\xi_1^2) \right].
\label{Sec3.1}
\end{eqnarray}
Evaluating the r.h.s.\ of Eq.~(\ref{Sec3.1}), we obtain
\begin{eqnarray}
N_{10} &=& 
\frac{17\pi^4}{720}
+ \frac{7}{2}\zeta(3) \ln 2 - 4  \Si_{1,3}(-1) 
\nonumber\\
&=& 
\frac{49\pi^4}{720} -  \frac{1}{2} b_4,
\label{res1}
\end{eqnarray}
where 
\begin{equation}
b_4 =-\frac{1}{3}(\pi^2-\ln^2 2)\ln^2 2+8\Li_4\left(\frac{1}{2}\right).
\label{b4}
\end{equation}
Equation~(\ref{res1}) coincides with the result obtained in
Ref.~\cite{Schroder}.

The evaluation of the $O(\epsilon)$ term $M_{10}$ is rather tedious and cannot
be described in this brief communication.
Here, we merely list the result, which reads
\begin{equation}
M_{10} = -\frac{149\pi^4}{180} \ln 2+ \frac{279}{8} \zeta(5)  - 7 b_5,
\label{M10}
\end{equation}
where 
\begin{equation}
b_5 =\frac{1}{45}(5\pi^2-3\ln^2 2)\ln^3 2+8\Li_5\left(\frac{1}{2}\right).
\label{b5}
\end{equation}
We note in passing that also the other master integrals presented in
Ref.~\cite{Schroder} can be written in a more compact form if the combinations
$b_4$ and $b_5$ of Eqs.~(\ref{b4}) and (\ref{b5}), respectively, are
introduced.

\section{Second example}
\label{sec:four}

Let us consider the second example of Ref.~\cite{CKMS}, which may be
graphically represented as
\begin{equation}
m^2
\raisebox{-0.9cm}{{
\includegraphics[height=2.0cm,bb=71 114 697 720]{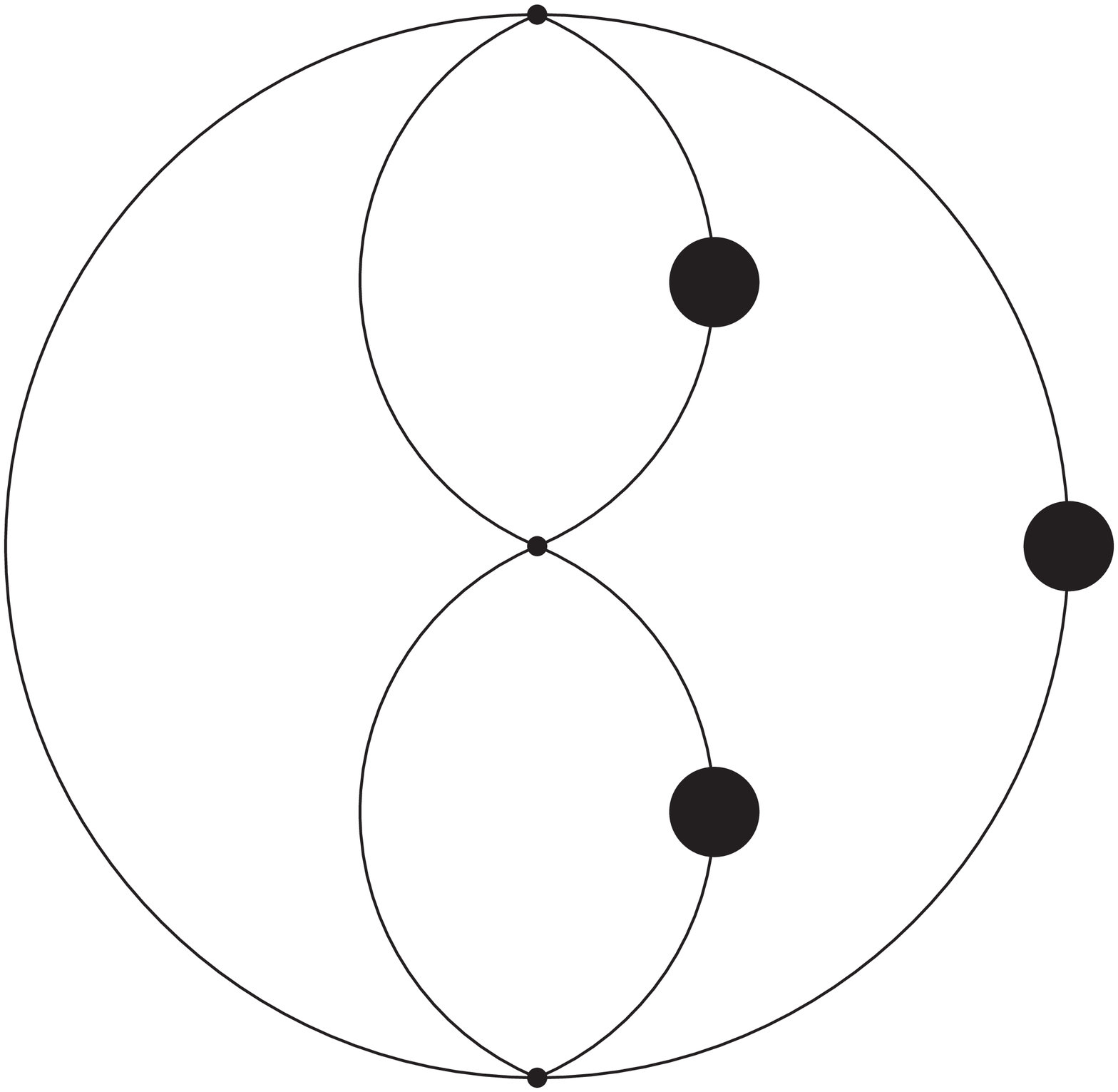}
}}
=\frac{\Gamma^4(1+\epsilon)}{(m^2)^{4\epsilon}}
N_{20}(\epsilon),
\label{NM2}
\end{equation}
where $N_{20}(\epsilon)=[N_{20}+M_{20}\epsilon+O(\epsilon^2)]/(1-\epsilon)$.
For $\epsilon=0$, we recover $N_{20}=N_{20}(0)$.
Applying Eq.~(\ref{eq:master}) with $\alpha_1=1$ and $\alpha_2=2$ to the two
interior massive loops and to the loops involving exterior propagators, we
easily obtain the following representation for $N_{20}(\epsilon)$:
\begin{equation}
N_{20}(\epsilon)=
\frac{(1-\epsilon)(m^2)^{4\epsilon}}{\Gamma(1+\epsilon)}
\int_0^1\frac{ds_1}{s_1\eta_1^{\epsilon}}
\int_0^1 \frac{ds_2}{s_2\eta_2^{\epsilon}}
\int_0^1 \frac{ds_3}{s_3\eta_3^{\epsilon}}
T_{M_1,M_2,M_3}(1+\epsilon,1+\epsilon,1+\epsilon).
\label{N20ep}
\end{equation}
Expanding the propagators $(k^2+M^2_2)^{-\alpha_2}$ and 
$(k^2+M^2_3)^{-\alpha_3}$ in\break
$T_{M_1,M_2,M_3}(\alpha_1,\alpha_2,\alpha_3)$ as
in Eq.~(\ref{exp}) and using Eq.~(\ref{eq:one}) for the tadpole
$T_{0,M_1}(0,\alpha_2)$, we obtain
\begin{eqnarray}
T_{M_1,M_2,M_3}(\alpha_1,\alpha_2,\alpha_3) &=&
\sum_{n=0}^{\infty} 
\frac{\Gamma(n+\alpha_2)}{n!\Gamma(\alpha_2)}
\sum_{l=0}^{\infty}
\frac{\Gamma(l+\alpha_3)}{l!\Gamma(\alpha_3)}\,
\frac{\Gamma(l+n+\overline{\alpha}-D/2)}{\Gamma(l+n+\overline{\alpha})}
\nonumber\\
&&{}\times
\frac{\mu_{12}^n\mu_{13}^l}{(M^2_1)^{n+l+\overline{\alpha}-D/2}}
\nonumber\\
&=& 
\frac{R(0,\overline{\alpha})}{(M_1^2)^{\overline{\alpha} -D/2}}
F_1(\overline{\alpha} -D/2,\alpha_2,\alpha_3;\overline{\alpha};\overline{x},
\mu_{13}/M^2_1),
\nonumber
\end{eqnarray}
where $F_1$ denotes an Appel hyper-geometric function \cite{Gradshteyn}.
Expanding the hyper-geometric function, we have
\begin{eqnarray}
\lefteqn{
T_{M_1,M_2,M_3}(1+a_1\epsilon,1+a_2\epsilon,1+a_3\epsilon)
=\frac{1}{(M_1^2)^{(\overline{a}+1)\epsilon}}\,
\frac{\Gamma(1+\epsilon)}{\mu_{32}(1-\epsilon)}}
\nonumber\\
&&{}\times\left\{ \frac{M_2^2}{\mu_{12}}\ln\frac{M_2^2}{M_1^2}
+ \epsilon \left[
- \frac{a_2+1}{2}\, \frac{M_2^2}{\mu_{12}}\ln^2\frac{M_2^2}{M_1^2}
+\left(\overline{a} \frac{M_1^2}{\mu_{12}} -a_1-a_3 \right)
\Li_2\left( \frac{\mu_{12}}{M_1^2}\right)
\right.\right. 
\nonumber\\
&&{}+\left.\left.
a_3\frac{M_1^2}{\mu_{12}}\Li_2\left(\frac{\mu_{13}}{M_1^2}\right)
+a_3\frac{M_2^2}{\mu_{12}} \left(\Li_2\left( \frac{\mu_{21}}{M_2^2}\right)-
\Li_2\left( \frac{\mu_{23}}{M_2^2}\right)\right)\right]
+ O(\epsilon^2)\right\}
\nonumber\\
&&{}+(2 \leftrightarrow 3).
\label{compli}
\end{eqnarray}

As in the previous section, we describe the evaluation of the $O(\epsilon^0)$
term in some detail, while we merely list the final result for the
$O(\epsilon)$ term for reasons of space.
At $O(\epsilon^0)$, the right-hand side of the Eq.~(\ref{N20ep}) contains the
tadpole $T_{M_1,M_2,M_3}(1,1,1)$, which can be calculated in an essentially
simpler way.
Indeed, using 
\begin{equation}
\frac{1}{(k^2+M^2_1)(k^2+M^2_2)}=\frac{1}{\mu_{21}}
\left(\frac{1}{k^2+M^2_1}-\frac{1}{k^2+M^2_2}\right)
\end{equation}
and Eq.~(\ref{eq:one}), we find
\begin{eqnarray}
T_{M_1,M_2}(1,1)&=&\frac{\Gamma(\epsilon-1)}{\mu_{21}}
\left[{(M^2_1)}^{1-\epsilon}-{(M^2_2)}^{1-\epsilon}\right]
\nonumber\\
&=&\frac{1}{\mu^{2\epsilon}}\, \frac{\Gamma(1+\epsilon)}{1-\epsilon}
\left(\frac{1}{\epsilon}+\frac{M^2_1}{\mu_{21}}\ln\frac{M^2_1}{\mu^2}\right)
+(M_1\leftrightarrow M_2)
\nonumber\\
&&{}+O(\epsilon),
\label{Sec4.0}\\
T_{M_1,M_2,M_3}(1,1,1)&=&\frac{1}{\mu_{32}}
\left[T_{M_1,M_2}(1,1)-T_{M_1,M_3}(1,1)\right]
\nonumber\\
&=&\frac{M^2_1}{\mu_{12}\mu_{13}}\ln\frac{M^2_1}{\mu^2}
+(M_1\leftrightarrow M_2)+(M_1\leftrightarrow M_3)
\nonumber\\
&&{}+O(\epsilon).
\label{Sec4.1}
\end{eqnarray}
Notice that the right-hand side of Eq.~(\ref{Sec4.1}) is $m^2$ and $\mu^2$
independent.
Moreover, it coincides with Eqs.~(\ref{N20ep}) and (\ref{compli}) at
$O(\epsilon^0)$.
Every term in Eqs.~(\ref{Sec4.0}) and (\ref{Sec4.1}) contains only one
logarithm depending on the variable $\eta_i$.
Thus, the integrals $\int^1_0 ds_i$ can be evaluated at the end of the
calculation (see Eq.~(\ref{Sec3.1}), for a similar procedure).
After some algebra, we obtain
\begin{equation}
N_{20}=\frac{9}{2}\zeta(3),
\end{equation}
which agrees with Ref.~\cite{Schroder}.

Lack of space prevents us from going into details with our derivation of the
$O(\epsilon)$ term $M_{20}$, and we merely list our result,
\begin{equation}
M_{20} = -\frac{4\pi^4}{15} - 27 \zeta(3) + 3 b_4,
\end{equation}
where $b_4$ is defined in Eq.~(\ref{b4}).

\section{Conclusion}
\label{sec:five}

We demonstrated the usefulness of a simple technique for the analytic
evaluation of four-loop tadpoles, which allows one to obtain the results by
integrating one-loop tadpoles with masses that depend on integration
variables.
By means of this method, we calculated two Feynman diagrams which were
presented in Ref.~\cite{CKMS} in numerical form.
Our results are in full agreement with the analytic results obtained just
recently in Ref.~\cite{Schroder}.

We note that the integral $N_{10}$ was found in Ref.~\cite{Schroder} with the
aid of the PSLQ program \cite{PSLQ}, which is able to reconstruct the
rational-number coefficients multiplying a given set of transcendental numbers
from a high-precision numerical result, i.e.\ the algebraic structure of the
result must be known a priori.
Such structures are often more complicated, and they are in general not known
from other considerations (see, for example, Ref.~\cite{FKalK}), so that the
PSLQ program cannot be applied.
The method presented here does not suffer from such a limitation and should
be applicable also for more complicated tadpoles 
depending on one \cite{Schroder} or two \cite{difmass} nonzero masses.
Moreover, it can be applied in combination with the PSLQ program.
In fact, by evaluating some part of a complicated Feynman diagram by means of
this method, the underlying set of transcendental numbers may emerge and can
then be injected into the reconstruction of the full analytic result with the
help of the PSLQ program.

\vspace{0.5cm}

\noindent{\bf Acknowledgements}

We are grateful to K.G. Chetyrkin for useful communications and for bringing
Ref.~\cite{Schroder} to our attention, and to O.L. Veretin for his interest in
this work, for his helpful comments, for checking some of our formulae, and
for technical advice related to the PSLQ program \cite{PSLQ}, which helped us
in checking the $O(\epsilon)$ terms of the four-loop tadpoles $N_{10}$ and
$N_{20}$.
A.V.K. was supported in part by the RFBR Foundation under Grant No.\
05-02-17645-a and the Heisenberg-Landau-Programm.
This work was supported in part by BMBF Grant No.\ 05 HT4GUA/4 and HGF Grant
No.\ NG-VH-008.

\end{document}